\documentclass[conference]{IEEEtran}
\IEEEoverridecommandlockouts
\usepackage{cite}
\usepackage{amsmath,amssymb,amsfonts}
\usepackage{subfigure}
\usepackage{comment}
\usepackage{algorithmic}
\usepackage{graphicx}
\usepackage{textcomp}
\usepackage{xcolor}
\def\BibTeX{{\rm B\kern-.05em{\sc i\kern-.025em b}\kern-.08em
    T\kern-.1667em\lower.7ex\hbox{E}\kern-.125emX}}
\begin{document}

\title{Evaluation of the Topological Agreement of Network Alignments}

\author{\IEEEauthorblockN{ Concettina Guerra}
\IEEEauthorblockA{\textit{College of Computing} \\
\textit{Georgia  Institute of Technology}\\
Atlanta, USA \\
guerra@gatech.edu}
\and
\IEEEauthorblockN{ Pietro Hiram Guzzi}
\IEEEauthorblockA{\textit{Dept Surgical and Medical Sciences} \\
\textit{Magna Graecia University}\\
Catanzaro, Italy \\
hguzzi@unicz.it}
}

\maketitle

\begin{abstract}
Aligning protein interaction networks (PPI) of two or more organisms consists of finding a mapping of the nodes (proteins) of the networks that captures important structural and functional associations (similarity). It is a well studied  but  difficult problem. It is provably NP-hard in some instances thus computationally very demanding. 
The problem comes in several versions: global versus local alignment; pairwise versus multiple alignment; one-to-one  versus many-to-many alignment.  Heuristics to address the various instances of the problem  abound  and they achieve some degree of success when their performance is measured in terms of node and/or edges conservation. However, as the evolutionary distance between the organisms being considered increases the results tend to degrade. 
Moreover, poor performance is achieved when the considered networks have remarkably different sizes in the number of nodes and/or edges.  Here we address the challenge of analyzing and comparing different approaches to global network alignment, when a one-to-one mapping is sought. We consider and propose various measures to evaluate the agreement between alignments obtained by existing approaches. We show that some such measures indicate an agreement that is often about the same than what  would be obtained by chance. That tends to occur even when the mappings exhibit a good performance based on standard measures.  
\end{abstract}

\begin{IEEEkeywords}
Network Alignment, Network Alignment Comparison, Graph Comparison
\end{IEEEkeywords}

\section{Introduction}

Interactions among proteins, as well as other biological molecules, are usually modelled using a formalism coming from graph theory \cite{Thula}. In such a scenario, biological entities are represented as nodes of a graph, while their interactions as  edges \cite{carrington2005models}. 
A PPI network is graph $G=(V,E)$ where $V$ is the set of nodes or proteins and $E$ the set of edges representing protein interactions.

The use of networks enables the use of the set of existing methodologies and algorithms for solving interesting and relevant biological problems. The comparison of networks has gained much attention in computational biology. Such a comparison is translated into a graph alignment problem \cite{Guzzi 2017}. There exist many variants of such a problem: pairwise or multiple, global or local, one-to-one versus many-to-many mappings. 

In this paper we consider global one-to-one alignments of two PPI networks obtained as a result of the application of different methods proposed in the literature with the goal of evaluating their agreement.
If $G_1=(V_1, E_1)$ and $G_2=(V_2, E_2)$ are the PPI networks of two organisms and $A$ and $B$ are two different  alignment methods
we denote by $f: G1 \rightarrow G_2$ and $g: G_1 \rightarrow G_2 $ the one-to-one mappings of $G_1$ into $G_2$ obtained by $A$ and $B$, respectively. Performance measures to assess the merits of a specific method exist  in the literature. They include the edge correctness  EC \cite{Kuchaie 2010}, the node correctness NC \cite{Patro2012}, and the S$^3$ score \cite{vijayan2014}, among others. Based on such measures, computed separately for each mapping, the relative performance of two methods can be established.

Two alignments may have similar performance in terms of EC or S$^3$, however they may exhibit remarkable differences in two important aspects: a) in the set of nodes of $G_2$ that are identified  as the corresponding of the nodes of $G_1$ and b) in the topology of the sub-networks of $G_2$ induced by the mappings $f$ and $g$.
Here we address the problem of quantitatively evaluating the agreement between the two alignments considering both aspects.
We investigate measures that are general in that they are suitable for comparing any two graphs as well as measures that are specific to this application.  

The first group  of measures evaluate the agreement on nodes; it includes the Jaccard index and the Cayley distance that .
Briefly, the Jaccard index  measures the overlap of  the subsets  of nodes of $G_2$ corresponding of the nodes of $G_1$ in $f$ and $g$. 
The Cayley distance is defined on permutations and counts the number of cycles or transpositions in the two permutations identified by the mappings.  
It exploits the  natural correspondence between alignments and permutations. 

The second group of measures evaluate the agreement on the topology. Three measures are computed on the sub-graphs $H$ and $K$  of $G_2$ induced  by two mappings: 1) the $\chi^2$ distance of the degree histograms to measure the similarity in the degree sequences of $H$ and $K$, 2) the graphlet distribution, a generalisation of the degree distribution, that measures
the number of nodes that are incident to graphlets  \cite{gdda}, and 3) the difference in the number and size of cliques.
To the best of our knowledge the measures 1) and 3) have never been used for comparing alignments.

As representative of the class of one-to-one alignment methods we select  SANA \cite{SANA 2017}  and MAGNA++ \cite{vijayan2014}  which, based on the standard performance measures of EC, NC and S$^3$,  achieve  results that in most cases outperform other existing methods.
We  show that  the mappings obtained by SANA and MAGNA++ on pairs of PPI networks of various organisms  have an agreement which is generally low and in some cases close to that of random mappings.

The paper is structured as follows: Section \ref{sec:alignment} introduces the network alignment problem and the algorithms used in this paper; Section \ref{sec:eval} presents the evaluation methodology; then Section \ref{sec:results} presents and discusses our results; finally, Section \ref{sec:conclusion} concludes the paper.

\section{The alignment problem}
\label{sec:alignment}
We consider two networks $G_1=(V_1,E_1)$ and  $G_2=(V_2, E_2)$. 
We denote by $|V1|$ and $|V_2|$ the number of nodes of the network $G_1$ and $G_2$ and assume that $|V_1|<= |V_2|$.
The problem of finding an alignment between $G_1$ and  $G_2$ is to find a mapping $f$:$V_1 \rightarrow V_2$ such that an objective function measuring the  quality of the alignment is maximised. There are some main instances of the alignment: local (LNA) and global (GNA), considering the approach, and pairwise and multiple, considering the number of the alignment networks.

\begin{figure*}[ht]
     \centering
     \begin{subfigure}
         \centering
         \includegraphics[width=3 in]{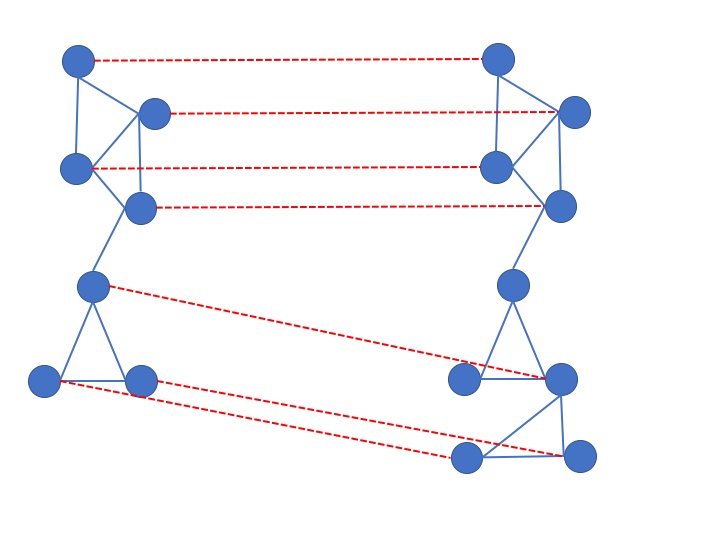}
         \label{fig:lna}
     \end{subfigure}
     \begin{subfigure}
         \centering
         \includegraphics[width=3 in]{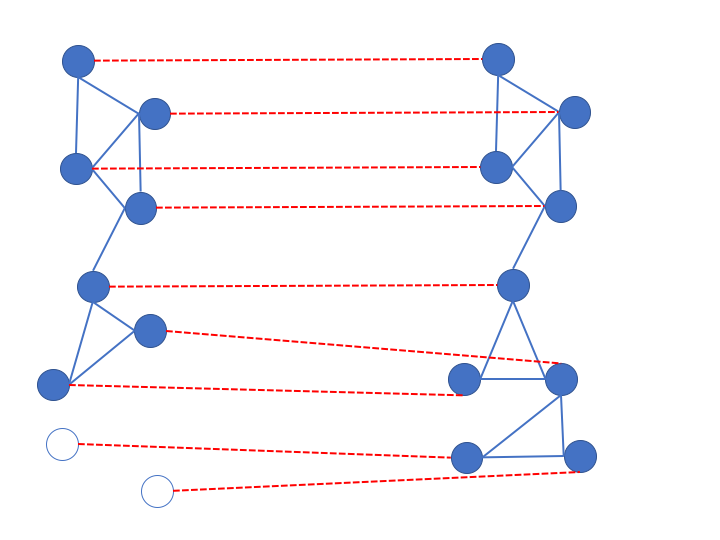}
         \label{fig:gna}
     \end{subfigure}
     \caption{LNA on the left vs GNA on the right.}
        \label{fig:gnavslna}
\end{figure*}

LNA \cite{cho2013m,mina2012i} searches for highly similar sub-graphs of the inputs that may represent conserved regions (e.g. conserved functional structures or protein complexes \cite{cannataro2010impreco}). Conversely, 
global network alignment GNA searches for the best mapping among all the nodes of the  input networks \cite{nassa2012} (adding dummy nodes when missing), which typically results in large but suboptimally conserved mapped subnetworks as depicted in Figure \ref{fig:gnavslna}.  From a biological perspective, LNA focuses on evolutionarily conserved building blocks of the cells. Instead, GNA searches for a single comprehensive mapping of the whole sets of protein interactions from different species. 
 The literature contains many GNA approaches.  The interested reader may find many details in a recent survey \cite{Guzzi 2017}. Some approaches, as IsoRank \cite{Isorank}, generate a many-to-many mapping that maps a subset of nodes of one network into a subset of nodes of the other, thus accounting for homology and orthology of proteins; other approaches generate a one-one-to mapping, that is is an injective function that takes a node of a network into a single node of the other.  
 
 Here we focus on global one-to-one mappings and select SANA \cite{SANA 2017} and  Magna++ \cite{vijayan2014} as representative of this class of methods since they outperform most of the existing methods \cite{SANA 2017}. 

SANA  is based on an optimization process that uses    
simulated annealing to optimize a given objective function.  This function can be any cost function, based either on node similarity or edge similarity. Simulated annealing tries many small changes, or moves, to the current solution  and based on the value of the recomputed objective function accepts or discards those moves. In fact, it accept also "bad" moves, based on a certain probability. Although this heuristic is typically expensive, it can run relatively fast  provided that the objective function is easy to compute.

MAGNA++  is an alignment method based on a genetic algorithm that simulates an evolutionary process to optimize both edge and node conservation.  The genetic algorithm requires an initial population of a given number of alignments. Such population  evolves over time in an iterative process that goes through several new generations.
During the evolution, the members of a population crossover with each other, and the child alignment  reflects each parent.  An alignment is represented by a permutation. MAGNA++ defines the crossover function of two permutations as the midpoint of the shortest path between them, where the distance between two permutations is the Cayley distance. 


The objective functions that SANA, MAGNA++ and other alignment methods seek to optimize include the  node conservation,  the edge correctness (EC) \cite{Kuchaie 2010}, the
Induced Conserved Structure (ICS) \cite{Patro2012}, and the score S$^3$ proposed in \cite{vijayan2014} or some combination of those.

For a given mapping $f$ of  $G_1$ into $G_2$, EC is the ratio of the number of edges conserved by $f$ to the number of edges $|E_1|$ in $G_1$: $EC(f)=|f(E1)/|E1|$. EC does not take into account the edges of the second network $G_2$. As such,  it does not penalize alignments of sparser network regions to denser ones.  

ICS \cite{Patro2012} is  the ratio of the number of edges conserved by $f$ to the number of edges in the sub-network of $G_2$ induced on the nodes in $G_2$ that are aligned to the nodes in $G_1$, denoted by $G_2(f(V_1))$. Thus, 
$ICS=\frac{|f(E_1)|}{E(G_2(f(V_1))}$

Finally, the score S$^3$ takes into account the edges of both $G_1$ and $G_2$: S$^3(f)=\frac{|f(E1)|}{|E1|+|E(G2(f(V1))|-|f(E1)|}$.
The above three scores differ only in the denominator.  The S$^3$ score is a trade-off between the other two measures since EC penalises misaligned edges in the smaller network while ICS penalises misaligned edges in the larger network.

The publicly available tools implementing  MAGNA++ and SANA allow to select the objective function to be optimized. In our comparative analysis we selected the S$^3$ score in both.

\section{Evaluation Methodology}
\label{sec:eval}

When evaluating the performance of a mapping, sometime one can rely on the   
the existence of a gold standard, i.e. a \textit{true} alignment. For global network alignments,  the node and edge correctness are based on the count of the nodes or edges that are \textit{correctly} mapped \cite{Milano 2017}. However, in most cases no such gold standard exists.  The measures below are used to address this challenge.

\subsection{Measure based on nodes}{\label{Sec:A}}
\bigskip\par\noindent
{\em Jaccard index}
The Jaccard index is a measure of the overlap of two sets. We use it to measure the overlap  of the sets of  nodes of a network that are identified by two different mappings as corresponding of the nodes of another network. This measure allows to gauge node-overlap
between the two graphs without node correspondence information.

Let $G^f_2$ and $G^g_2$ the subsets of nodes of $G_2$ that the nodes of $G_1$ map into in the mapping $f$ and $g$, respectively. More precisely, $G^f_2 = \{ y \in  G_2 : y=f(x)$ for  some $x \in  G_1 \}$; 
$G^g_2 = \{ y \in G_2 : y=g(x)\,$  for some $x \in  G_1 \}$.  It is $|G^f_2|=|G^g_2|=|V_1| \leq |V_2|$ since the mappings are one-to-one. 
The Jaccard index is defined as :
 \medskip\par\noindent
\centerline{$Jaccard_{index} = \frac{|G^f_2  \cap G^g_2|}{|G^f_2 \cup G^g_2|}$}
\medskip\par\noindent
that is the Jaccard index is the size of the intersection of two sets divided by the size of their union.

For reference, we compared the Jaccard index of the mappings of  real networks  with that of random mappings. 
One set of experiments was performed on random networks obtained by degree-preserving randomization.  More in details, each real PPI network  was rewired by randomly selecting  edges $(a,b)$ and $(c,d)$ and replacing them by the edges (a,c) and (b,d); this process was  repeated multiple  times (in our experiments 10 times the size of the graph). The results were two  random networks $G_1'=(V_1, E_1')$ and $G_2'=(V_2, E_2')$ with the same set of nodes and degree sequence as the original ones.  The Jaccard index was computed on the mappings from $G_1'$ into $G_2'$ generated by SANA and MAGNA++.

The second set of experiments involved random subsets of nodes of the second network $G_2$.  Precisely,  we  generated pairs of sets  containing $|V_1|$ nodes randomly selected out of the $|V_2|$ nodes of $G_2$. Two such subsets can be thought of as containing nodes of $G_2$ corresponding to the nodes  of $G_1$ in two arbitrary mappings. Thus this randomization, unlike the previous one, does not preserve any feature of the original graph $G_1$. 
\bigskip\par\noindent
{\em Cayley distance}

One simple way to assess the agreement in two mappings $f$and $g$ is  to determine the number of nodes $x \in G_1$ for which  $f(x)=g(x)$. This is more specific than just determining whether a node $y$ of $G_2$ is in  both $f(V1)$ and $g(V_1)$, i.e.  $y=f(x)=g(z)$, as in the Jaccard index defined above. Clearly, due to the many homologous  and orthologous proteins present in PPI networks, one would not expect a large agreement in that measure. 
A more general measure, the Cayley distance, relaxes this property. 
An interesting case occurs when a subset of nodes of the  graph $G_1$ is  mapped by $f$ and $g$  into the same subset of nodes of $G_2$ although the individual nodes are mapped in different nodes of the subset.  That is the case of a clique of say $k$ nodes that is mapped into the a clique of $G_2$ by $f$ and $g$ but the nodes are traded with each other, as in isomorphic graphs. 
The Cayley distance, defined on permutations,  captures this type of behavior.

A mapping $f: G_1 \rightarrow G_2$  can be represented by a permutation of the nodes of $G_2$. 
We denote by $p$ and $q$ the permutations that are associated to $f$ and $g$, respectively. Both $p$ and $q$ have size $|V_1|$ and are incomplete if $|V_2| > |V_1|$. 
\par\noindent

The Cayley distance  of two permutations $p$ and $q$ 
is the minimum number of transpositions necessary to transform $p$ in $q$. A transposition is the exchange of two elements in a permutation. 
Cayley \cite{Diaconis} proved that this distance is equal to $n-c$ where $c$ is the number of cycles in $p^{-1}q$ and $p^{-1}$ is the inverse of permutation $p$. 
A cycle, also called orbit \cite{Comtet}, is a subset of elements of a permutation that trade places with one another.
Thus,   1-cycles (cycles of length 1) correspond to fixed elements in the permutations and therefore originate from two mappings $f$ and $g$ that map an element of $G_1$ into the same element of $G_2$. 
Next,  an $r$-cycle identifies $r$ elements of $G_2$ that are associated to the same set of $r$ elements of $G_1$ although in a different order. 
As an example, an r-cycle occurs when  $r$ nodes form a clique in both networks and the mappings exchange the order of such nodes. 

Distances on permutations do not easily extend to incomplete permutations and generally such extensions  are very expensive to compute. 
However,  simple closed form solutions exist for some problems \cite{Zhou 2018}. The extension of the Cayley distance   was derived  in \cite{Critchlow} and is given by the simple relation:
$C^*(p,q)=n-c^*$   
where $c^*$ is the number of cycles in $p^{-1}q$ which consist of integers less than or equal to $|V_1|$. 

In the following, when comparing two mappings, we report the value $c*$, i.e. the number of cycles in the mappings. Thus $c*$ is a measure of similarity.

 \subsection{Measures based on network topology}

There are endless ways of designing similarity measures to compare graphs based on topology. Two  recent surveys review and evaluate the ability of several measures in capturing important properties of the  graphs \cite{Tantardini 2019, Wills 2020}. Measures based on some form of graph isomorphism are expensive to compute for real graphs; others, such as feature-based measures,  are more practical while maintaining a good level of accuracy. Which measure is the best choice depends on the specific application being considered.  

The comparison of two mappings $f$ and $g$ can be formulated as the comparison of the two sub-graphs $H$ and $K$ of $G_2$ induced by $f$ and $g$, for which the existing measures of graph similarity can be used. 

Here we evaluate the agreement of two mappings in degree distribution and in the distribution of specific types of sub-graphs, such as cliques or graphlets. This latter has been extensively investigated; proposed measures include  Relative Graphlets Frequency Distance  \cite{Przulj 2004}, the graphlet degree distribution agreement measure  \cite{Przulj 2007}, and  
Graphlets Correlation Distance  \cite{Yave 2014}. 
\bigskip\par\noindent
{\em Degree distribution} 
We analyze the degree sequences of the two  sub-networks $H$ and $K$ of $G_2$ induced by $f$ and $g$ and compute their histogram. 
The distance of the two histograms, measured by the  $\chi ^2$ function, gives an estimate of the lack of agreement of $f$ and $g$. 
The $\chi ^2$ distance is defined as:
\begin{equation}\label{chi}
\chi ^2 (P, Q) = 1/2 \sum_i \frac{(Pi - Qi)^2}{Pi + Qi}
\end{equation}
where $P_i$ and $Q_i$ are the values of bins $i$ in the histograms $P$ and $Q$ and give the number of nodes of degree $i$. 

This measure takes into account the fact that  the difference between large bins may less important than
the difference between small bins and should be given less weight. This may be appropriate for PPI networks  with large number of nodes of degree 0 or 1 and few high degree nodes reflecting the power law distribution of such networks. 

For an estimate of its significance, the $\chi ^2$  distance of mappings of real networks is  evaluated against that of  random  networks  obtained by degree-preserving randomization, as described in section \ref{Sec:A}.
\bigskip\par\noindent
{\em Graphlet degree distribution}
The Graphlet Degree Distribution Agreement (GDDA) \cite{gdda} is a generalisation of the degree distribution and measures
the number of nodes that are incident to graphlets. Graphlets, as defined in \cite{Przulj 2004}, are small sub-graphs of a
network that are connected and non-isomorphic. The GDDA compares the similarity
of distributions of each automorphism orbit within two networks. It
considers 73 degree distributions of graphlets with 2-5 nodes.
Considering all the possible combinations of the 73 graphlet degree
distributions the GDDA produces a measure of similarity (or agreement)
between two networks whose values range from 0 to 1, where 1 means
identity.

\bigskip\par\noindent
{\em Distribution of  Cliques} 
Here we take a somewhat different view and consider as sub-graphs of interest only cliques and determine the difference in the number of cliques in the two graphs.
Let 
$CN_k(H) $ and $CN_k(K)$ be the numbers of $k$-cliques in the sub-graph $H$ and $K$ induced by $f$ and $g$. 
The {\em clique distribution distance} CDD is the sum over all $k$ of the absolute values of the difference of $CN_k(H) $ and $CN_k(K)$   normalized by the sum of the same two values. 
\begin{equation}
CDD=  \sum_k  \frac{|CN_k(H) - CN_k(k)|}{CN_k(H) + CN_k(k)}.
\end{equation}



\section{Results on the comparison of alignments}
\label{sec:results}

In this section we evaluate the agreement of  alignment methods SANA and MAGNA++ on several pairs of PPI networks. The two methods allow to select the objective function to  be optimized among the  global topological scores of EC or ISC and S$^3$ or to select a function based on node functional similarity based on  GO terms. We experimented with  the two methods using the S$^3$ score as the objective function and as inputs various PPI networks of bacteria and eukaryotes.

\subsection{Datasets}

We used eight public available networks  whose dimensions are reported in in Table \ref{tab:networksize}. The networks represent protein interactions of eight organisms: Campilobacter jejuni \textit{(Cjeuni)}, Mesorhizobium loti \textit{(meso)}, Synechocystis sp. \textit{(syne)},Escherichia coli \textit{(ecoli)} ,C. Elegans \textit{(Celeg)}, S. Cerevisiae \textit{(Scere)}, D. Melanogaster \textit{(Dmela)}, Homo Sapiens \textit{(HSapi)}.

\begin{table}[ht]
  \caption{Sizes of the considered PPI networks}
  \centering
\begin{tabular}{|l|c|c|}
\hline
Network & n. of nodes & n. of edges\\
\hline
Campylobacter jejuni (cjejuni)& 1095 & 2988 \\ 
\hline
Mesorhizobium loti (meso) & 1803 & 3094 \\
\hline
Synechocystis sp. (syne) & 1908 & 3102 \\
\hline
Escherichia coli (ecoli) & 1941 & 3989 \\
\hline
C. Elegans (Celeg) & 2912 & 5298 \\
\hline
S. Cerevisiae (Scere) & 5831 & 77149 \\
\hline
D. Melanogaster (Dmela) & 7937 & 34753\\ 
\hline
Homo Sapiens (HSapi) &13212 & 110495\\
\hline
\end{tabular}
\label{tab:networksize}
\end{table}

\subsection{Results based on the agreement on nodes}

We computed the Jaccard index and the number of Cayley cycles of each pair of PPI networks using Magna++ and SANA. The results are reported in Tables  \ref{tab:Jaccard} and \ref{tab:Cayley}.

 \begin{table*}[hbt!]  
\centering
\caption{ The table reports the Jaccard index  of pairs of mappings of bacteria, eukaryotes, and random networks.}
\begin{tabular}{| l | c |c | c| c | c| c|}
\hline
\multicolumn{7}{|c|}{\bf Agreement based on nodes: Jaccard Index} \\
\hline
\multicolumn{7}{|c|}{\it  Real networks - Bacteria} \\
\hline
& cjejuni & cjejuni  & cjejuni&  meso & meso & syne\\
& ecoli &  syne& meso  & ecoli &   syne  & ecoli \\
\hline
Jaccard index& 0.42 & 0.46 & 0.46 &  0.89 & 0.92  & 0.98 \\
\hline
\multicolumn{7}{|c|}{\it  Random networks (by rewiring)} \\
\hline
avg Jaccard index & 0.43 & 0.46 & 0.44 & 0.9 & 0.92 & 0.97\\
\hline
\multicolumn{7}{|c|}{\it  Random subsets of nodes of the second network} \\
\hline
avg Jaccard index &  0.393 & 0.398 & 0.43 & 0.87 & 0.88& 0.98\\
\hline
\end{tabular}
\label{tab:Jaccard}
\end{table*}

\begin{table*}[hbt!]
\centering
\begin{tabular}{| l | c |c | c| c | c| c|}
\hline
\multicolumn{7}{|c|}{\it  Real networks - Eukaryotes} \\
\hline
& Celeg & Celeg & Celeg & Scere & Scere & Dmela \\
& HSapi   & Dmela & Scere & Dmela &  HSapi & HSapi  \\
\hline
Jaccard index & 0.08 & 0.14 & 0.24   & 0.48 &  0.39 & 0.34\\
\hline
\multicolumn{7}{|c|}{\it  Random subsets of nodes of the second network} \\
\hline
avg Jaccard index &  0.05 &  .12 & 0.22 & 0.33 &  0.28 &  0.3\\
\hline
\end{tabular}
\label{tab:RandomJaccard}
\end{table*}

\begin{table*}[hbt!]
\centering
\caption{ The table reports the number of Cayley cycles  of pairs of mappings of bacteria and eukaryotes}
\begin{tabular}{| l | c |c | c| c | c| c|}
\hline
\multicolumn{7}{|c|}{\bf Agreement based on nodes: n. of Cayley cycles c*} \\
\hline
\multicolumn{7}{|c|}{\it  Bacteria} \\
\hline
& cjejuni & cjejuni  & cjejuni&  meso & meso & syne\\
& ecoli &  syne& meso  & ecoli &   syne  & ecoli \\
\hline
Cayley cycles ($c*$) & 0& 0 & 3 & 7  & 8 & 9 \\
\hline
\multicolumn{7}{|c|}{\it Eukaryotes} \\
\hline
& Celeg & Celeg & Celeg & Scere & Scere & Dmela \\
& HSapi   & Dmela & Scere & Dmela &  HSapi & HSapi  \\
\hline
\hline
Cayley cycles ($c*$) & 0 &2 & 1 &1   &  7 & 0 \\
\hline
\hline
\end{tabular}

\label{tab:Cayley}
\end{table*}

We first observe that the Jaccard index is high and close to 1 when the networks have approximately the same size, as in the case of syne and ecoli. This is to be expected since in such cases  the number of possible ways of mapping the nodes of $G_1$ into a subset of nodes of $G_2$  is relatively small (recall the mappings are injective and $|G_1| \leq |G_2|$) . 

On the other hand, the Jaccard index is low when the networks differ significantly in the number of nodes.  Consider, for instance, the pair of PPI networks  cjejuni and ecoli.  MAGNA++ and SANA identify subsets of nodes of ecoli that are remarkably different: the sets $G^f_2$ and $G^g_2$ have size 1095 and their intersection has size 652  corresponding to a Jaccard index of  0.41.
 
Interestingly, those values are about the same as those obtained by the intersection of two randomly selected subsets of $G_2$ of size $|G_1|$. The values in  table \ref{tab:Jaccard} (lower part) are the average Jaccard indexes  over 10,000 pairs of random subsets.
For the case of random networks obtained by degree-preserving rewiring the average Jaccard index is even lower indicating an better agreement between random networks than between real networks. 

The Jaccard index of some eukaryotes is lower than that of bacteria;  it is 0.14 the pair Scere and Dmela and 0.24 for Cele and Dmela. Such values are close to those obtained by intersecting  random subsets of nodes of Dmela. Note that for eukaryotes we did not compute the mappings of randomized graphs with rewiring since MAGNA++ tends to be time consuming on such large graphs.

It turns out that the Jaccard index is inversely correlated to the difference of the number of nodes  of the two networks. In fact, for all bacterial networks the correlation between the Jaccard index and the difference in size is -0.99. While it is obvious that the agreement is high when the difference in size of the two networks is small it is interesting that it decreases remarkably when the difference in size increases.  

As for the number of Cayley cycles, shown in table \ref{tab:Cayley}, this is generally very small, further stressing the lack of consensus of the mappings. For instance, no cycles are identified for the mappings of cjejuni into ecoli. That is,  not only $f(x) \neq g(x)$ for all  nodes x in cjejuni but also  there is no group of nodes in one mapping that trade places (order) in the other mapping. We notice that, as for the Jaccard index, this measure indicates a better agreement when the networks are close in size. 

\subsection{Results on the agreement in topology}
\subsubsection*{Comparison based on degree distribution}
We computed the $\chi^2$  distance in degree distribution of the two induced sub-graphs $H$ and $K$ according to  the expression in (\ref{chi}). Note that we did not include in the sum the bins corresponding to degrees 0 and 1. The $\chi^2$ values  are  shown in table \ref{tab:chi-square} for all pairs of considered bacteria. They are also shown for pairs of random networks obtained from $G_1$ and $G_2$ by degree preserving randomization. As seen from the table, the  $\chi^2$ distances of the actual networks are sometimes higher than those of the corresponding random networks indicating a better agreement for random networks. This tends to happen  when $G_2$ has high variance in the degree sequence, as for instance in meso. When  $G_2$ is characterized by a small range of degrees, as in ecoli, approximately the same distance is found for random and real networks.
\begin{table*}[!ht] 
\centering
\caption{For each pair of PPI networks the table reports the $\chi^2$ distance of the degree histograms of the sub-graphs induced by the mappings by SANA and MAGNA++.}
\begin{tabular}{| l | c |c | c| c | c| c|}
\hline
\multicolumn{7}{|c|}{\bf Distance of alignments} \\
\multicolumn{7}{|c|}{\bf based on degree distribution} \\
\hline
\multicolumn{7}{|c|}{\it  Real networks } \\
\hline
& cjejuni & cjejuni  & cjejuni&  meso & meso & syne\\
& ecoli &  syne& meso  & ecoli &   syne  & ecoli \\
\hline
$\chi^2$ distance & 41.7 & 42.3 & 65.8 & 62.  & 39.9  & 23.6 \\
of induced sub-graphs &&&&&&\\
\hline
\multicolumn{7}{|c|}{\it  Random networks } \\
\hline
avg $\chi^2$ distance & 41.8  &   35.9&  38.4 & 33.6 & 31.9& 23.5\\
\hline
\end{tabular}
\label{tab:chi-square}
\end{table*}
\subsubsection*{Comparison using  GDDA}
The GDDA values expressing the agreement in the graphlet degree distribution of the induced sub-graphs $H$ and $K$ of $G_2$ are in table \ref{tab:gdda} along with  the GDDA values of pairs of random  networks.

From the table we observe that there is little  agreement in the induced sub-graphs $H$ an $K$ as measured by GDDA  (always less than 0.2) and it does not vary much across all pairs of organisms. For random networks obtained by rewiring $G1$ and $G2$, the agreement is even lower, suggesting that the mappings preserve the topological property under consideration for actual networks better than for random networks, although in both cases the agreement is negligible.

\begin{table*}[!ht] 
    \centering
      \caption{For each pair of PPI networks the table reports the GDDA Agreement of the sub-graphs induced by the mappings by SANA and MAGNA++}
    \begin{tabular}{|c|c|c|c|c|c|} \hline
    \multicolumn{6}{|c|}{\bf Agreement of alignments in graphlet distribution} \\
\multicolumn{6}{|c|}{\bf GDD-Agreement with Arithmetic Mean (GDDA)} \\ \hline 
\multicolumn{6}{|c|}{\it  Real networks } \\ \hline 
\multicolumn{6}{|c|}{\it  Sub-graphs $H$ and $K$ of $G_2$ induced by the mappings } \\ \hline

 cjejuni-ecoli&	cjejuni-syne&	cjejuni-meso&	meso-ecoli &	meso-syne &	syne-ecoli \\ \hline 
     
  0,13 &	0,11 &	0,19 &	0,16 &	0,15 &	0,13 \\ \hline
  \hline
  \multicolumn{6}{|c|}{\it Random networks}\\
  \hline
  \multicolumn{6}{|c|}{\it Sub-graphs $H$ and $K$ of $G'_2$ induced by the mappings} \\ \hline 
   0,09 &	0,08 &	0,05 &	0,05 &	0,10 &	0,03 \\ \hline
    \end{tabular}
  
    \label{tab:gdda}
\end{table*}

\subsubsection*{Comparison based on the distribution of cliques}
\begin{table*}[!htp] 
\centering
\caption{For each pair of PPI networks the table reports the number of k-cliques in the  sub-graphs induced by SANA and MAGNA++}
\begin{tabular}{|p{3cm} | c |c | c| c | c| c|c|c| c|c| }
\hline
\multicolumn{11}{|c|}{\bf Number of cliques in the PPI networks } \\

\hline
 & {\bf 3-cliques} & {\bf 4-cliques} &  {\bf 5-cliques} & {\bf 6-cliques} & {\bf7-cliques} & {\bf 8-cliques} & {\bf 9-cliques} & {\bf 10-cliques}&   {\bf 11-cliques}& {\bf 12-cliques}\\
\hline
cjejuni & 713 & 185 & 37 & 4 & -& -& -& -& -& -\\
\hline
meso & 200 & 1 & -& -& -& -& -& -& -& -\\
\hline
syne &55 &- & -& -& -& -& -& -& -& -\\
\hline
ecoli &4721 & 6672 & 9559 & 9475 &8149 &  2552 &1301 &516 & 132 & 16 \\
\hline
\hline
\multicolumn{11}{|c|}{\bf Number of cliques in the induced subgraphs } \\
\hline
\multicolumn{11}{|c|}{\it  \large cjejuni - ecoli} \\
\hline
sub-graph induced by SANA & 2272 & 3641 & 5010 & 4227 &2677  & 623 & 151 & 24 & 2& - \\
\hline
sub-graph induced by MAGNA++ & 2290 &  3640& 4813 &    4141& 2703 & 708 & 254 & 67& 8& -\\

\hline
\multicolumn{11}{|c|}{\it  \large  cjejuni - syne} \\
\hline
\hline
sub-graph induced by SANA & 41 & - &-  &-  &-  & - & - & - & -& - \\
\hline
sub-graph induced by MAGNA++ & 33 &  -& - &    -& - & - &- &- & -& -\\
\hline
\hline
\multicolumn{11}{|c|}{\it  \large  cjejuni - meso} \\
\hline
sub-graph induced by SANA & 115 & - &-  &-  &-  & - & - & - & -& - \\
\hline
sub-graph induced by MAGNA++ & 93 &  1& - &    -& - & - &- &- & -& -\\
\hline
\hline
\multicolumn{11}{|c|}{\it  \large  meso - ecoli} \\
\hline
sub-graph induced by  SANA & 1330 & 581 & 213 & 45 & 5 & - & - & - & -& - \\
\hline
sub-graph induced by  MAGNA++ & 2797 & 3538 & 4132 &  3204 &  1684  &  472 & 125 & 21& -& -\\
\hline
\hline
\multicolumn{11}{|c|}{\it  \large  meso - syne} \\
\hline
sub-graph induced by SANA & 33 & - &-  &-  &-  & - & - & - & -& - \\
\hline
sub-graph induced by MAGNA++ & 41 &  -& - &    -& - & - &- table
&- & -& -\\
\hline
\hline
\multicolumn{11}{|c|}{\it  \large  syne - ecoli} \\
\hline
sub-graph induced by SANA &  3431 & 3791 &4299  &3559  &2611  & 764 & 357 & 112 & 16& - \\
\hline
sub-graph induced by MAGNA++ & 4068 &  5373& 7404 & 7241   & 6431 & 2072 & 1169& 495 & 130& 16\\
\hline
\end{tabular}
\medskip
\begin{tabular}{| l | c |c | c| c | c| c|}
\hline
\multicolumn{7}{|c|}{}\\
\multicolumn{7}{|c|}{\bf Distance of the distributions of k-cliques } \\
\hline
& cjejuni & cjejuni  & cjejuni&  meso & meso & syne\\
& ecoli &  syne& meso  & ecoli &   syne  & ecoli \\
\hline
CDD  &1.4 &0.1 & 1.9 & 6.9 & 0.1 & 4.6 \\ 
\hline
\end{tabular}
\label{tab:cliques}
\end{table*}

The results of the analysis on the distribution of k-cliques of the sub-graphs $H$ and $K$ induced by the mappings of SANA and MAGNA++ are reported in table \ref{tab:cliques}.
The table gives also the number of k-cliques of the input PPI networks of the bacteria. 

We first observe the tendency of both methods to map nodes of $G_1$  into cliques of $G_2$ even when no corresponding cliques exist in $G_1$. For instance,  the number of cliques in the  sub-graphs of ecoli induced by the mapping of syne into ecoli  is of the order of thousands in both SANA and MAGNA++ even though  syne has only few cliques of size 3. This is explained by the fact that  the two alignment methods optimize the measure S$^3$ which favors nodes with high degree in order to obtain a large number of conserved edges.  
However, for the pair syne-ecoli as well as the pair meso-ecoli there is a noticeable difference in the number and distribution of cliques in SANA and MAGNA++ as expressed by the CCD distance (see the lower part of  table \ref{tab:cliques}).
This is another way in which the outputs of two methods differ significantly.
By contrast, the distance CDD is very low between mappings of networks that are both relatively sparse, as in syne and meso.

In conclusion the CCD measure indicates a lack of agreement in the way SANA and MAGNA++ map nodes of a network into another especially when the second network is denser than the first.

We remark that we did not run experiments on the eukaryotes using  CDD because of the large size of the inputs leading to long running times of MAGNA++.

\section{Conclusions}
\label{sec:conclusion}

We have analyzed the agreement on nodes and topology of two popular network alignment methods, SANA and MAGNA++, on  pairs of bacteria and eukaryotes. We observed a remarkable  difference in the alignments even when the two methods optimized the same score ($S^3$ in our experiments).

The similarity in terms of the nodes was measured using the Jaccard index and the Cayley number of cycles. They  evaluate the overlap of the sets of nodes of $G_2$ that the nodes of $G_1$ map into.  For several pairs of organisms, such measures indicate an agreement close to what would be obtained with random networks generated by degree-preserving randomization or by random selection of subsets of nodes of the second network. In a way, this result is consistent with the result in \cite{Kazemi} that showed that the node similarity metric of IsoRank \cite{Isorank} when only the structural properties are used is a function of the nodes  degrees only  
and does not depend on the actual edge set of the two networks.  In other words, similarity is invariant to any network degree-preserving rewiring. 

As for topological agreement of the sub-graphs induced by the mappings, we expressed that in terms of the distributions of their degrees ($\chi^2$),  graphlets (GDDA) and cliques (CCD). For all such measures, the analysis based on topology did not reveal a much different scenario than that based on nodes, especially when considering the measure CCD.
Unlike the distance  GDDA that is very low but almost uniform across the different pairs of organisms, the distance  CDD shows high variation across the organisms, being very low when the alignments map a network into a denser one. 

Although such lack of agreement may appear a weakness, on the other hand it may be exploited in more than one way. First, it allows the exploration of different mappings resulting in a better insight into the biological relatedness of two organisms. Second it can lead to better alignments when addressing the multiple alignments problem that seeks the alignment of the PPI networks of multiple species. This study is a preliminary step to find a methodology for building a multiple alignment by composing many pairwise alignments obtained by different aligners.
\color{black}



\section*{Funding}
Guerra was partially supported by US-Israel Binational Grant (BSF) n. 2018141.


\end{document}